\documentclass[12pt]{article}

\usepackage{amsmath,amssymb,array}
\usepackage{graphicx}
\usepackage{comment,color}

\newcommand{\bs}{\boldsymbol}

\allowdisplaybreaks

\def\e{{\rm e}}
\def\c{\cos}

\def\s{{\rm s}}
\def\T{{\rm T}}
\def\H{{\rm H}}
\def\SM{{\rm SM}}
\def\tr {{\rm Tr}}
\def\nt{\notag}
\def\gm{\gamma^{\mu}}		
			
\def\dy{\int_{-\pi R}^{\pi R}\!\!\!\!\!\!\!\!{\rm d}y~}%
\def\sig{i\sigma^2}

\def\tt{\times10}

\makeatletter
    
    \@addtoreset{equation}{section}
\makeatother

\begin{document}
\setlength{\baselineskip}{18pt}
\begin{titlepage}

\begin{flushright}
KOBE-TH-10-01
\end{flushright}
\vspace{1.0cm}
\begin{center}
{\Large\bf Flavor Mixing in Gauge-Higgs Unification} 
\end{center}
\vspace{25mm}

\begin{center}
{\large
Yuki Adachi, 
%
Nobuaki Kurahashi, 
%
C. S. Lim
%
%
and Nobuhito Maru$^*$
}
\end{center}
\vspace{1cm}
\centerline{{\it Department of Physics, Kobe University,
Kobe 657-8501, Japan}}

\centerline{{\it
$^*$Department of Physics, Chuo University, 
Tokyo 112-8551, Japan
}}
%
%
\vspace{2cm}
\centerline{\large\bf Abstract}
\vspace{0.5cm}
We discuss flavor mixing and resultant flavor changing neutral current processes
in the $SU(3) \otimes SU(3)_\text{color}$ gauge-Higgs unification scenario.
To achieve flavor violation is a challenging issue in the scenario,
since the Yukawa couplings are originally higher dimensional gauge interactions.
We argue that the presence of $Z_2$-odd bulk masses of fermions
plays a crucial role as the new source of flavor violation.
Although introducing brane-localized mass terms in addition to the bulk masses
is necessary to realize flavor mixing,
if the bulk masses were universal among generations,
the flavor mixing and flavor changing neutral current processes are known to disappear.
We also discuss whether natural flavor conservation is realized in the scenario.
It is shown that
the new source of flavor violation leads to flavor changing neutral current processes
at the tree level due to the exchange of non-zero Kaluza-Klein gauge bosons.
As a typical example we calculate the rate of $K^0 - \bar{K}^0$ mixing
due to the non-zero Kaluza-Klein gluon exchange at the tree level.
The obtained result for the mass difference of neutral kaon is suppressed
by the inverse powers of the compactification scale.
By comparing our prediction with the data
we obtain the lower bound of the compactification scale
as a function of one unfixed parameter of the theory,
which is of ${\cal O}(10)$ TeV, except for some extreme cases.
We argue that the reason to get such rather mild lower bound is the presence
of ``GIM-like" mechanism, which is a genuine feature of GHU scenario.

\end{titlepage}




\newpage

\section{Introduction}

In spite of the great success of the standard model, the mechanism of spontaneous gauge symmetry breaking is still not conclusive. 
Higgs particle, responsible for the spontaneous breaking seems to have various theoretical problems, 
such as the hierarchy problem and the presence of many arbitrary coupling constants of its interactions.  

Gauge-Higgs unification (GHU) \cite{Manton, Fairlie, Hosotani} is one of the attractive scenarios 
solving the hierarchy problem without invoking supersymmetry. In this scenario, 
Higgs doublet in the Standard Model (SM) is identified with the extra spatial components of the higher dimensional gauge fields. 
Remarkable feature is that the quantum correction to Higgs mass is insensitive to the cutoff scale of the theory and calculable 
regardless of the non-renormalizability of higher dimensional gauge theory, which is guaranteed by the higher dimensional gauge invariance.  
This fact has opened up a new avenue to the solution of the hierarchy problem \cite{HIL}. 
The finiteness of the Higgs mass has been studied and verified in various models 
and types of compactification at one-loop level\footnote{For the case of gravity-gauge-Higgs unification, 
see \cite{HLM}.} \cite{ABQ} and even at the two loop level \cite{MY}. 

The fact that the Higgs is a part of gauge fields indicates that the Higgs interactions are basically governed by gauge principle. 
Thus, the scenario may also shed some light on the long standing arbitrariness problem in the interactions. 

To see whether the scenario is viable, it will be of crucial importance to address the following questions. 

\noindent (1) Does the scenario have characteristic and generic predictions on the observables subject to the precision tests ?  

\noindent (2) Though there is a hope that the problem of the arbitrariness of Higgs interactions may be solved, 
how are the variety of fermion masses and flavor mixing realized ?  

\noindent (3) In view of the fact that Higgs interactions are basically gauge interactions with real gauge coupling constants, 
how is the CP violation realized ? 

Let us note that the problems (2) and (3) are also shared by superstring theories, where the low energy effective theory, 
i.e. the point particle limit, of the open string sector is (10-dimensional) super Yang-Mills theories, which can be regarded as a sort of GHU.  
 
Concerning the issue (1), it will be desirable to find out finite (UV-insensitive) and calculable observables subject to the precision tests, 
although the theory is non-renormalizable and observables are very UV-sensitive in general. 
Works from such viewpoint already have been done in the literature, \cite{LM}, \cite{LHC}, \cite{ALM1}. 
The issue (3) has been addressed in recent papers \cite{ALM3}, \cite{LMN}, 
where CP violation is claimed to be achieved ``spontaneously" by the VEV of the Higgs field 
or by the geometry of the compactified extra space.  

In this paper, we focus on the remaining issue (2) concerning flavor physics in the GHU scenario. 
Let us note that it is highly non-trivial problem to account for the variety of fermion masses and flavor mixings in this scenario, 
since the gauge interactions are universal for all generations of matter fields, while (global) symmetry among generations, say flavor symmetry, 
has to be broken (flavor violation) to distinguish each flavor and to get their mixings.  

A genuine feature of the higher dimensional theories with orbifold compactification is 
that the gauge invariant bulk mass terms for fermions, generically written as $M \epsilon (y) \bar{\psi} \psi$ with $\epsilon (y)$ 
being the sign function of extra space coordinate $y$, are allowed. 
The bulk masses may be different depending on each generation and can be an important new source of the flavor violation. 
The presence of the mass terms causes the localization of Weyl fermions in two different fixed points of the orbifold 
depending on their chiralities and the Yukawa coupling obtained by the overlap integral over $y$ of the mode functions of Weyl fermions with different chiralities is suppressed by a factor $2\pi RM\e^{-\pi RM} \ (R: {\rm the size of the  extra space})$. Thus observed hierarchical small fermion masses can be achieved without fine tuning thanks to the exponential suppression factor $\e^{-\pi RM}$.  

One may expect that the bulk masses need not to be diagonal in the base of generation and lead to the flavor mixing. 
Unfortunately, it is not the case: for each representation $R$ of gauge group, $\psi (R)$, 
the bulk mass terms are generically written as $M(R)_{ij} \epsilon (y) \bar\psi(R)_{i} \psi (R)_j \ (i,j: \mbox{generation index})$ 
by use of hermitian matrix $M(R)_{ij}$, which can be diagonalized by a suitable unitary transformation keeping the kinetic terms untouched. 
So the fermion mass eigenstates are essentially equal to gauge (weak) eigenstates and the flavor mixing may not occur in the bulk space. 
We are thus led to introduce a brane localized interaction to achieve the flavor mixings as was proposed in \cite{BN}. 
The brane localized interaction results in a discrepancy between mass eigenstates and gauge eigenstates. 
It is interesting to note that though the brane localized interaction contains theoretically unfixed parameters 
behaving as the source of flavor mixing, the bulk masses still play important roles.  
Namely as we will see below, in the limit of universal bulk masses among generations the flavor mixing exactly disappears, 
basically because in this limit there is no way to distinguish each generation of bulk fermions. Thus the interplay between brane localized interaction and bulk mass is crucial to get flavor mixing. This is a remarkable feature of the GHU scenario, 
not shared by, e.g., the scenario of universal extra dimension 
where flavor mixing may be caused by Yukawa couplings just as in the standard model, irrespectively of bulk masses.  

Knowing that the flavor mixing is realized it will be important to discuss flavor changing neutral current (FCNC) processes, 
which have been playing a crucial role for the check of viability of various new physics models, 
as is well-known in the case of supersymmetric models. 
A central issue is whether ``natural flavor conservation" is realized, 
i.e. whether FCNC processes at tree level are ``naturally" forbidden in the GHU scenario. 
In ordinary four-dimensional framework, 
there exists a useful condition discussed by Glashow and Weinberg \cite{GW} to ensure the natural flavor conservation: 
{\it fermions with the same electric charge and the same chirality should possess the same quantum numbers, 
such as the third component of weak isospin $I_{3}$}. 
We find if we restrict ourselves to the pure zero mode sector with light masses, 
the condition is satisfied and there is no FCNC process at the tree level, which may be a natural consequence since our model reduces to the standard model at low energies. 
However, as a new feature of higher dimensional model, 
in the low energy processes of zero mode fermions due to the exchange of non-zero Kaluza-Klein (KK) modes of gauge bosons 
the FCNC processes are known to be possible already at the tree level, 
even though the amplitudes are suppressed by the inverse powers of compactification scale due to the decoupling of heavy gauge bosons. 
This fact is caused by the generation-dependent bulk masses of fermions, 
which is a new source of flavor violation, not considered in the argument of Glashow-Weinberg \cite{GW}. 
Such bulk masses cause non-trivial profiles in the extra space for the mode functions of fermions 
and together with the oscillatory mode functions of non-zero KK gauge bosons the overlap integrals of mode functions 
lead to non-universal (generation-dependent) couplings of the massive gauge bosons. 
Let us note in the case of zero mode gauge bosons, 
the couplings are inevitably universal simultaneously with the normalization of kinetic terms of fermions. 

In our paper, such general argument is confirmed in a typical concrete example of FCNC process, $K^{0} - \bar{K}^{0}$ mixing. 
What we calculate is the dominant contribution to the process, the tree diagram with the exchange of non-zero KK gluons. 
Comparing the obtained finite contribution to the mixing with the experimental data, 
we put the lower bound on the compactification scale.   

This paper is organized as follows. 
In the next section, 
we introduce our model and discuss the mass eigenvalues and mode functions of fermions. 
In section 3, we perform a general analysis how the flavor mixing is realized in our model. 
In section 4, as an application of our results in section 3, 
we calculate the mass difference of neutral kaons caused by the $K^0-\bar{K}^0$ mixing via non-zero KK gluon exchange at the tree level.  
We also obtain the lower bound for the compactification scale, by comparing the obtained result with the experimental data. 
Our conclusions are given in section 5.  



\section{The Model}
We consider a five dimensional $SU(3)\otimes SU(3)_\text{color}$ GHU model 
compactified on an orbifold $S^1/Z_2$ with a radius $R$ of $S^1$. The $SU(3)$ unifies the 
electro-weak interactions $SU(2) \otimes U(1)$.  
As matter fields, 
we introduce $n$ generations of bulk fermion in the fundamental representation 
and the (complex conjugate of) second-rank symmetric tensor representation of $SU(3)$ gauge group, 
$\psi^i({\bf 3}) =  Q_{3}^i \oplus d^i$ and $\psi^i(\bar {\bf 6}) = \Sigma^i \oplus Q_6^i \oplus u^i \ (i = 1,2,\ldots, n)$, 
which contain ordinary quarks of the standard model in the zero mode sector, 
i.e. a pair of $SU(2)$ doublet, $Q_{3}^i$ and $Q_6^i$, and $SU(2)$ singlets, $d^i$ and $u^i$. 
$\psi^i(\bar {\bf 6})$ also contain $SU(2)$ triplet exotic states $\Sigma^i$ \cite{BN}.  

The bulk Lagrangian is given by 
\begin{align}
 \mathcal{L}
=&
 -\frac{1}{2}\tr (F_{MN}F^{MN})
 -\frac{1}{2}\tr (G_{MN}G^{MN})
 \nt\\
&
 +\bar{\psi}^i({\bf 3})(i\!\not \!\!D -M^i\epsilon(y))\psi^i({\bf 3})
 +\bar{\psi}^i({\bar{\bf 6}})(i\!\not \!\!D -M^i\epsilon(y))\psi^i({\bar{\bf 6}})
\end{align}
where 
\begin{align}
 F_{MN}
&=
 \partial_MA_N-\partial_NA_M -ig[A_M,A_N],
 \\
 G_{MN}
&=
 \partial_MG_N-\partial_NG_M -ig_\s[G_M,G_N],
 \\
 \not \!\! D 
&=
 \Gamma^M (\partial_M -igA_M-ig_\s G_M). 
\end{align}
The gauge fields $A_M$ and $G_M$ are written in a matrix form, 
e.g. $A_M = A_M^{a} \frac{\lambda^{a}}{2}$ in terms of Gell-Mann matrices $\lambda^{a}$. 
It should be understood that $A_M$ in the covariant derivative $D_M = \partial_M -igA_M-ig_\s G_M$ acts properly 
depending on the representations of the fermions and $G_M$ acts on the color indices. 
$M,N=0,1,2,3,5$ and the five dimensional gamma matrices are $\Gamma^M=(\gamma^{\mu},i\gamma^{5})$ ($\mu=0,1,2,3$). 
$g$ and $g_\s$ are five-dimensional gauge coupling constants of $SU(3)$ and $SU(3)_\text{color}$, respectively. 
$M^i$ are $Z_2$-odd generation dependent bulk masses of the fermions with $\epsilon (y) = 1$ and $-1$ for $y > 0$ and $y <0$, respectively.

The periodic boundary condition is imposed along $S^1$ and $Z_2$ parity
assignments are taken for gauge fields as
\begin{align}
A_{\mu}
=&
 \left(
 \begin{array}{ccc}
  (+,+)&(+,+)&(-,-)\\
  (+,+)&(+,+)&(-,-)\\
  (-,-)&(-,-)&(+,+)
 \end{array}
 \right),
 ~~
 A_y
=
 \left(
 \begin{array}{ccc}
  (-,-)&(-,-)&(+,+)\\
  (-,-)&(-,-)&(+,+)\\
  (+,+)&(+,+)&(-,-)
 \end{array}
 \right),
 \nt\\
 G_{\mu}
=&
 \left(
 \begin{array}{ccc}
  (+,+)&(+,+)&(+,+)\\
  (+,+)&(+,+)&(+,+)\\
  (+,+)&(+,+)&(+,+)
 \end{array}
 \right),
 ~~
 G_y
=
 \left(
 \begin{array}{ccc}
  (-,-)&(-,-)&(-,-)\\
  (-,-)&(-,-)&(-,-)\\
  (-,-)&(-,-)&(-,-)
 \end{array}
 \right),  
\label{BC}
\end{align}
where (+,+) etc. stand for  $Z_2$ parities at fixed points $y=0, \pi R$.
We can see that the gauge symmetry $SU(3)$ is explicitly broken to $SU(2) \times U(1)$ by the boundary conditions. 
The fermions are assigned the following $Z_2$ parities with all colors having the same parity: 
\begin{align} 
 \Psi^i({\bf 3})
&=
 \big(Q_{3L}^i(+,+)+Q_{3R}^i(-,-)\big) \oplus \big(d_L^i(-,-)+d_R^i(+,+)\big), 
 \nt\\
 \Psi^i(\bar{\bf 6})
&=
 \big(\Sigma^i_L(-,-) + \Sigma^i_R(+,+)\big) \oplus \big(Q_{6L}^i(+,+)+Q_{6R}^i(-,-)\big) \oplus \big(u^i_L(-,-)+u^i_R(+,+)\big).  
\label{BC2}
\end{align}
Thus a chiral theory is realized in the zero mode sector by $Z_2$ orbifolding.  


Let us derive fermion mass eigenvalues and mode functions necessary for the argument of flavor mixing. 

The fundamental representation $\psi^i({\bf 3})$ is expanded by an ortho-normal set of mode functions as follows:
\begin{equation}
 \psi^i({\bf 3})
=
 \left[
 \begin{array}{c}
 Q_{3L}^if_L^i(y)+\sum\limits_{n=1}^{\infty}\left\{Q_{3L}^{i(n)}f_L^{i(n)}(y) +Q_{3R}^{i(n)}S_n (y) \right\}\\[10pt]
 d_{R}^if_R^i(y)+\sum\limits_{n=1}^{\infty}\left\{d_{R}^{i(n)}f_R^{i(n)}(y) +d_{L}^{i(n)}S_n (y) \right\}\\
 \end{array}
 \right], 
\end{equation}
where 
the mode functions are given in \cite{ALM3}: $S_n (y) = \frac{1}{\sqrt{\pi R}} \sin (M_n y)$ and 
\begin{align}
 f^{i}_L(y)&=\sqrt{\frac{M^i}{1-\e ^{-2\pi RM^i}}}\e^{-M^i|y|}, \quad
 f^{i}_R(y)= \sqrt{\frac{M^i}{\e^{2\pi RM^i}-1}}\e^{M^i|y|},\\
 f_L^{i(n)}(y)
&=
 \frac1{\sqrt{\pi R}}\frac{M_n}{m^i_n}
  \left[\c \left( M_ny \right) - \frac{M^i}{M_n} \epsilon(y)\sin \left(M_ny \right) \right],
 \\
 f_R^{i(n)}(y)
&=
 \frac1{\sqrt{\pi R}}\frac{M_n}{m^i_n}
  \left[ \c \left( M_ny \right) + \frac{M^i}{M_n}\epsilon(y)\sin \left( M_ny\right) \right],
\end{align}
with $M_n=n/R$, $m_n^i=\sqrt{(M^i)^2+M_n^2}$. 
The mode functions in the first line are those for the zero modes,  
and those in the second and the third lines are for non-zero KK modes.  
We can see that before the spontaneous electroweak symmetry breaking the fermion mass terms are diagonalized 
by use of these mode functions:
\begin{equation}
\begin{aligned}
\dy
  \bar{\psi}^i({\bf 3})\big(i\Gamma^5\partial_5 -M^i \epsilon(y)\big)\psi^i({\bf 3})
&=
 \sum_{n=1}^{\infty} m_n^i\!
 \left[ \bar Q^{i(n)}_3Q_3^{i(n)} - \bar{d}^{i(n)} d^{^i(n)} \right]
\\
&\to 
 -\sum_{n=1}^{\infty}m_n^i\!
 \left[\bar Q^{i(n)}_3Q_3^{i(n)} + \bar{d}^{i(n)} d^{i(n)} \right]
\end{aligned}
\end{equation}
In the second line, a chiral rotation 
$Q_3^{i(n)}\to \e ^{i\frac{\pi}{2} \gamma_{5}}Q_3^{i(n)}$ is performed.

The second-rank symmetric tensor representation $\bar {\bf 6}$ in a matrix form can be decomposed into 
three different $SU(2) \times U(1)$ representations as follows: 
\begin{equation}
\begin{aligned}
 \psi^i\big(\bar {\bf 6}\big)
=&
 \left[
 \begin{array}{c|c}
 \big(i\sigma^2\big)\Sigma^i\big(i\sigma^2\big)^{\!\T}&\frac1{\sqrt2}\big(i\sigma^2\big)Q_6^i
 \\[5pt]\hline
 \frac1{\sqrt2}Q_6^{i\T}\big(i\sigma^2\big)^{\!\T} & u^i
 \end{array}
 \right],
\end{aligned}
\end{equation}
where $\sig$ denotes an $SU(2)$ invariant anti-symmetric tensor
$\big(\sig\big)^{\alpha\beta}\!=\epsilon^{\alpha\beta}$.
Each component is expanded by the same mode functions as in the fundamental representation:
\begin{equation}
\begin{aligned}
 \Sigma^i
&=
 \Sigma_R^if_R^i(y)
 +\sum_{n=1}^{\infty}
 \left[
 \Sigma_R^{i(n)}f_{R}^{i(n)}(y) +\Sigma_L^{i(n)}S_n (y) 
 \right],
 \\
 Q_6^i
&=
 Q_{6L}^if_L^i(y)
 +\sum_{n=1}^{\infty}\left[Q_{6L}^{i(n)}f_L^{i(n)}(y)+Q_{6R}^{i(n)}S_n (y) \right],
\\
 u^i
&=
 u_{R}^if_R^i(y)  
 +\sum_{n=1}^{\infty}\left[u_{R}^{i(n)}f_R^{i(n)}(y)+u_{L}^{i(n)}f_L^{i(n)}(y)\right].
\end{aligned}
\end{equation}
The mass terms of $\psi\big(\bar {\bf 6}\big)$ are also diagonalized, ignoring the VEV of $A_y$:  
\begin{align}
 {\rm Tr}\,
 \bar \psi^i\big(\bar {\bf 6}\big)\!
 \big(i\Gamma^5 \partial_5-M^i\epsilon(y)\big)
 \psi^i (\bar { \bf 6})
&=
 -\!\sum_{n=1}^\infty m_n^i\!
 \left[{\rm Tr}\,\bar \Sigma^{i(n)}\Sigma^{i(n)}
 +\bar Q_{6}^{i(n)}Q_{6}^{i(n)}
 +\bar u^{i(n)} u^{i(n)}\right]
\end{align}
where a chiral rotation $Q_6^{i(n)}\to \e^{i\frac{\pi}{2}\gamma^5}Q_6^{i(n)}$ is performed.

We notice that there are two massless quark doublets $Q_{3L}$ and $Q_{6L}$ per generation in this model. 
In a simplified one generation case, for instance, 
one of two independent linear combinations of these doublets should correspond to the ordinary quark doublet of the standard model, 
but another one is an exotic state. 
Moreover, we have an exotic fermion $\Sigma_R$. 
We therefore introduce brane localized four dimensional Weyl spinors $Q_R$ and $\chi_L$ 
to form $SU(2) \times U(1)$ invariant brane localized Dirac mass terms in order to 
remove these exotic massless fermions from the low-energy effective theory.
\begin{equation}
\begin{aligned}
 \mathcal{L}_\text{BM}
=&
 \dy \sqrt{2\pi R}\,\delta(y)\bar Q_R^i(x)
 \Big[\eta_{ij}Q_{3L}^j(x,y)+\lambda_{ij}Q_{6L}^j(x,y)\Big]
\\
&
 +\dy\sqrt{2\pi R}\,m_{\rm BM}\delta(y-\pi R)\bar \Sigma_R^i(x,y)\chi_{L}^i(x)
 +({\rm h.c.})
 \end{aligned}
\end{equation}
where $Q_R$ and $\chi_L$ behave as doublet and triplet of $SU(2)$ respectively. 
The $n \times n$ matrices $\eta_{ij}, \lambda_{ij}$ and $m_{\rm BM}$ are mass parameters. 
These brane localized mass terms are introduced at opposite fixed points 
such that $Q_R(\chi_L)$ couples to $Q_{3,6L}(\Sigma_R)$ localized on the brane at $y=0~(y=\pi R)$. 
Let us note that the matrices $\eta_{ij}, \lambda_{ij}$ can be non-diagonal, which becomes the origin of the flavor mixing \cite{BN}.   

Some comments on this model are in order. 
The predicted Weinberg angle of this model is not realistic, $\sin^2 \theta_W = 3/4$. 
Possible way to cure the problem is to introduce an extra $U(1)$ or 
the brane localized gauge kinetic term \cite{SSS}. 
However, the wrong Weinberg angle is irrelevant to our argument,   
since our interest is in the flavor mixing and resultant 
$K^0-\bar K^0$ mixing via KK gluon exchange in the QCD sector, whose amplitude is independent of the Weinberg angle.  

Second, the bulk masses of fermions are generation-dependent, but are taken as common for both $\psi^i({\bf 3})$ and $\psi^i(\bar {\bf 6})$. 
In general, the bulk masses of each representation are arbitrary and there is no reason to take such a choice. 
It would be a natural choice if we have some Grand Unified Theory (GUT) 
where the $\bf 3$ and $\bar {\bf 6}$ representations are embedded into a single representation of 
the GUT gauge group.\footnote{For instance, if we consider the following gauge symmetry breaking pattern 
$Sp(8) \to Sp(6) \times SU(2) \to SU(3) \times U(1) \times SU(2)$, 
then we find that ${\bf 3}$ and $\bar{{\bf 6}}$ of $SU(3)$ can be embedded into the adjoint representation ${\bf 36}$ of $Sp(8)$ \cite{Slansky}. 
This is because the adjoint representation is decomposed as follows; 
${\bf 36} \to ({\bf 1}, {\bf 3}) \oplus ({\bf 21}, {\bf 1}) \oplus ({\bf 6}, {\bf 2}) 
\to ({\bf 1}, {\bf 3}) \oplus ({\bf 1} \oplus {\bf 6} \oplus \bar{{\bf 6}} \oplus {\bf 8}, {\bf 1}) \oplus ({\bf 3} \oplus \bar{{\bf 3}}, {\bf 2})$.}

\section{Flavor mixing}

As we noticed in the introduction, the flavor mixing does not occur in the bulk 
since the weak eigenstate is essentially equal to the mass eigenstate due to the fact that Yukawa coupling is originated from the gauge coupling. 
Then the argument in the previous section suggests 
that the brane localized mass terms for the doubled doublets $Q_{3L}$ and $Q_{6L}$ may lead to the flavor mixing. 
We now confirm the expectation and discuss how the flavor mixing is realized in this model. 
Let us focus on the sector of quark doublets and singlets, which contain fermion zero modes.  
First, we identify the quark doublet of the standard model by diagonalizing the relevant brane localized mass term, 
\begin{align}
 \dy \sqrt{2\pi R}\,\delta(y)\bar Q_R(x)
 \big(\eta, \lambda\big)\!\!
 \left(\begin{array}{c}Q_{3L}(x,y)\\Q_{6L}(x,y)\end{array}\right)
&\supset
 \sqrt{2\pi R}\,\bar Q_R(x)
 \big(\eta f_L(0),\lambda f_L(0)\big)\!\!
 \left(\begin{array}{c}Q_{3L}(x)\\Q_{6L}(x)\end{array}\right)
 \nt\\
&=
 \sqrt{2\pi R}\, \bar Q_R'(x)
 \big(m_\text{diag}, \bs0_{n\times n}\big)\!\!
 \left(\begin{array}{c}Q_{\H L}(x)\!\!\\Q_{\SM L}(x)\!\!\end{array}\right)
\label{diagonal} 
\end{align}
where 
\begin{align}
& \left(\begin{array}{cc}U_1 & U_3 \\U_2&U_4\end{array}\right)\!\!
 \left(\begin{array}{c}Q_{\H L}(x)\\Q_{\SM L}(x)\end{array}\right)
 =
 \left(\begin{array}{c}Q_{3L}(x)\\Q_{6L}(x)\end{array}\right), \quad 
U^{\bar Q} Q_R(x)=Q_R'(x),
\\
& U^{\bar Q}\big(\eta f_L(0),\lambda f_L(0)\big)\! \left(\begin{array}{cc} U_1 & U_3 \\ U_2 & U_4 \end{array} \right) 
=
 (m_\text{diag}, \bs0_{n\times n}).
\end{align}
In eq. (\ref{diagonal}), $\eta f_L(0)$ is an abbreviation of a $n \times n$ matrix whose $(i, j)$ element is given by $\eta_{ij}f_L^{j} (0)$, for instance.
$U_3$, $U_4$ are $n \times n$ matrices which compose a $2n \times 2n$ unitary matrix diagonalizing the brane localized mass matrix. 
The eigenstate $Q_\H$ becomes massive and decouples from the low energy processes, 
while $Q_\SM$ remains massless at this stage and therefore is identified with the quark doublet of the standard model. 
$U_1,\cdots,U_4$ satisfy the following unitarity condition:
\begin{align}
\label{unitary_cond}
 \left\{
 \begin{aligned}
 &U_1^\dag U_1+U_2^\dag U_2=U_3^\dag U_3+U_4^\dag U_4={\bf 1}_{n\times n}\\
 &U_1^\dag U_3+U_2^\dag U_4={\bf 0}_{n\times n}
 \end{aligned}
 \right., \quad 
 \left\{
 \begin{aligned}
 &U_1 U_1^\dag+U_3U_3^\dag=U_2U_2^\dag +U_4 U_4^\dag={\bf 1}_{n\times n}\\
 &U_1 U_2^\dag+U_3 U_4^\dag={\bf 0}_{n\times n}
 \end{aligned}
 \right.. 
\end{align}
After this identification, we find Yukawa couplings are read off from the higher dimensional gauge interaction of $A_y$, 
whose zero mode is the Higgs field $H(x)$:  
\begin{align}
 &  \dy\!\!
	\left[
	-\frac g2\bar\psi^i({\bf 3}) A_y^a \lambda^a \Gamma^y \psi^i({\bf 3})
	+g{\rm Tr}
	\Big\{
	\bar\psi^i({\bar{\bf 6}})A_y^a(\lambda^a)^\ast\Gamma^y\psi^i({\bar{\bf 6}})
	\Big\}
	\right]\notag\\
 \supset&
	\dy\!\!\left\{
	-g\bar Q^{i}_{3L}(x,y) H(x,y) d^{i}_R (x,y)
	-\sqrt2g\bar Q^i_{6L}(x,y)i\sigma^2H^*(x,y)u^{i}_R(x,y)+({\rm h.c.})
	\right\}\notag\\
\supset&
	-\!g_4\!
	\left\{
	\big\langle H^{\dagger} \big\rangle\bar d_R^{i}(x)
	I_{RL}^{i(00)}U_3^{ij}Q_{\SM L}^{j}(x)
	+\sqrt2\big\langle H^t \big\rangle \sig
	\bar u_R^{i}(x)I_{RL}^{i(00)}U_4^{ij}Q_{\SM L}^{j}(x)
	\right\}
	+({\rm h.c.})
\end{align}
where $g_4\equiv \frac g{\sqrt{2\pi R}}$ and
\begin{equation}
   I_{RL}^{i(00)}
 = \dy f_L^if_R^i
 = \frac{\pi RM^i}{\sinh(\pi RM^i)}\ ,
\end{equation}
which behaves as $2\pi RM^i\e^{-\pi RM^i}$ for $\pi RM^i \gg 1$,
thus realizing the hierarchical small quark masses without fine tuning of $M^i$.
We thus know that the matrices of Yukawa coupling $g_4Y_u$ and $g_4Y_d$ are given as 
\begin{equation} 
\label{Yukawa coupling} 
	g_4Y_u = \sqrt2g_4I_{RL}^{(00)} U_4\ ,\qquad
	g_4Y_d = g_4I_{RL}^{(00)} U_3\ ,
\end{equation}
where the matrix $I_{RL}^{(00)}$ has elements $\big(I_{RL}^{(00)}\big)_{ij} = \delta_{ij} I_{RL}^{i(00)}$. 
These matrices are diagonalized by bi-unitary transformations as in the Standard Model
and Cabibbo-Kobayashi-Maskawa matrix is defined in a usual way.
\begin{equation}
\label{cond_U3,U4}
\left\{
\begin{aligned}
&\hat Y_d=\text{diag}(\hat m_d,\cdots)=
V_{dR}^\dag Y_d V_{dL},
\\
&\hat Y_u=\text{diag}(\hat m_u,\cdots)=
V_{uR}^\dag Y_u V_{uL},
\\
&V_\text{CKM}\equiv V_{dL}^\dag V_{uL},
\end{aligned}
\right.
\end{equation}
where all the quark masses are normalized by the W-boson mass as $\hat m_f =\frac{m_f}{m_W}$.
A remarkable point is that the Yukawa couplings $g_4Y_u$ and $g_4Y_d$ are mutually related by the unitarity condition eq.\,(\ref{unitary_cond}), 
while those in the Standard Model are completely independent. 
Thus if we set bulk masses of fermion to be universal among generations, i.e. $M^1=M^2=M^3=\cdots=M^n$, 
then $I_{RL}^{(00)}$ is proportional to the unit matrix. 
In such a case, $Y_u^\dag Y_u \propto U_4^{\dagger} U_4$ and $Y_d^\dag Y_d \propto U_3^{\dagger} U_3$ can be simultaneously diagonalized 
because of the unitarity condition eq.\,(\ref{unitary_cond}). 
This means that the flavor mixing disappears in the limit of universal bulk masses, as was expected in the introduction. 
In reality, off course the bulk masses should be different to explain the variety of quark masses and therefore the flavor mixing should be realized.  

For the illustrative purpose to confirm the mechanism of flavor mixing, let us work in a framework of two generations. 
We will see how the realistic quark masses and mixing are reproduced. 
The argument here will be useful also for the calculation in the next section.  
For simplicity, we ignore CP violation and assume that $U_3$, $U_4$ are real.  
By noting that an arbitrary $2\times2$ matrix can be written in a form $O_1M_{\rm diag}O_2$
in terms of 2 orthogonal matrices $O_{1,2}$ and a diagonal matrix $M_{\rm diag}$
and by use of unitarity condition (\ref{unitary_cond}),
$U_3^\dag U_3+U_4^\dag U_4 = {\bf 1}_{2\times 2}$,
the matrices relevant for the Yukawa couplings are known to be parameterized as 
\begin{align} 
\label{parametrization}
 &  U_4
  = \left(
	\begin{array}{cc}
	 \cos \theta' & -\sin \theta' \\
	 \sin \theta' & \cos \theta' \\
	\end{array}
	\right)\!\!
	\left(
	\begin{array}{cc}
	 a & 0 \\
	 0 & b \\
	\end{array}
	\right), \quad 
	U_3
  = \left(
	\begin{array}{cc}
	 \cos \theta & -\sin \theta \\
	 \sin \theta & \cos \theta \\
	\end{array}
	\right)\!\!
	\left(
	\begin{array}{cc}
	 \sqrt{1-a^2} & 0 \\
	 0 & \sqrt{1-b^2} \\
	\end{array}
	\right), \\ 
	\label{parametrization2} 
 &  I_{RL}^{(00)}=
	\left(
	\begin{array}{cc}
	 c & 0 \\
	 0 & d \\
	\end{array}
	\right). 
	\end{align}
The most general forms of $U_3$ and $U_4$ have a common orthogonal matrix
multiplied from the right, being consistent with (\ref{unitary_cond}).
The matrix can be absorbed into a rotation
among degenerate doublets of the standard model $Q_{\SM L}$ and has no physical meaning.
In another word, the common orthogonal matrix can be absorbed into $V_{dL}, \ V_{uL}$
without changing $V_{\text{CKM}}$.
Thus, without loss of generality we can adopt the parametrization (\ref{parametrization}). 

Physical observables $\hat m_u, \hat m_c, \hat m_d, \hat m_s$ 
and the Cabibbo angle $\theta_{c}$ are written in terms of $a,b,c,d$ and $\theta, \theta'$. 
Namely trivial relations
\begin{align}
 &  {\rm det} \big(\hat{Y}_d^\dag \hat{Y}_d\big) = \hat m_d^{2} \hat m_s^{2}\ ,\quad
	{\rm det} \big(\hat{Y}_u^\dag \hat{Y}_u \big) = \hat m_u^{2} \hat m_c^{2}\ ,\\
 &  {\rm Tr} \big(\hat{Y}_d^\dag \hat{Y}_d \big) = \hat{m}_d^2 + \hat{m}_s^2\ ,\quad
	{\rm Tr} \big(\hat{Y}_u^\dag \hat{Y}_u\big) = \hat{m}_u^2 + \hat{m}_c^2
\end{align}
provide through eqs.
(\ref{Yukawa coupling}), (\ref{cond_U3,U4}),
(\ref{parametrization}), (\ref{parametrization2}) with
\begin{align}
	\hat{m}_d^2 \hat{m}_s^2
 &= \big(1-a^{2}\big)\!\big(1-b^{2}\big) c^{2} d^{2}\ ,
	\label{cond1} \\
	\hat{m}_d^2 + \hat{m}_s^2
 &= \big(1-a^2\big)c^2+\big(1-b^2\big)d^2
	+\big(a^2 - b^2\big)\!\big(c^2 - d^2\big) \sin^{2} \theta\ ,
	\label{cond2} \\ 
	\hat{m}_u^2 \hat{m}_c^2
 &= 4a^{2} b^{2} c^{2} d^{2}\ ,
	\label{cond3} \\
	\hat{m}_u^2 + \hat{m}_c^2
 &= 2\left\{a^2c^2+b^2d^2-\big(a^2-b^2\big)\!\big(c^2-d^2\big)\sin^2\theta'\right\}\ .
	\label{cond4} 
\end{align}
We also note that the $\theta_c$ is given as 
\begin{align} 
&\tan 2\theta_c
    =\frac{\tan2\theta_{dL}-\tan2\theta_{uL}}{1+\tan2\theta_{dL}\tan2\theta_{uL}},
\label{Cabibbo} \\ 
&\tan 2\theta_{dL}
     =\frac{2\sqrt{(1-a^2)(1-b^2)}(d^2-c^2)\sin\theta\cos\theta}
      {(1-a^2)(c^2\cos^2\theta+d^2\sin^2\theta)-(1-b^2)(c^2\sin^2\theta+d^2\cos^2\theta)}, 
\label{Cabibbo1} \\   
&\tan2\theta_{uL}
     =\frac{2ab(d^2-c^2)\sin\theta'\cos\theta'}
      {a^2(c^2\cos^2\theta'+d^2\sin^2\theta')-b^2(c^2\sin^2\theta'+d^2\cos^2\theta')}, 
\label{Cabibbo2}  
\end{align}
where angles $\theta_{dL}, \ \theta_{uL}$ are angles parameterizing $V_{dL}, \ V_{uL}$, respectively. 
Let us note 5 physical observables are written in terms of 6 parameters, $a,b,c,d$ and $\theta, \theta'$. 
So our theory has 1 degree of freedom, which cannot be determined by the observables. Let us choose $\theta'$ as the unfixed parameter. Then once we choose the value of $\theta'$, other 5 parameters can be completely fixed by the observables, by solving eqs. (\ref{cond1}), (\ref{cond2}), (\ref{cond3}), (\ref{cond4}), (\ref{Cabibbo}), (\ref{Cabibbo1}) and (\ref{Cabibbo2}) numerically for $a,b,c,d$ and $\theta$. The result is shown in Table 1. 

\begin{figure}[h]
\centering
\setlength{\extrarowheight}{1pt}
\begin{tabular}{|c||c|c|c|c|c|}
 \hline
 $\sin\theta'$ & $a^2$ & $b^2$ & $c^2$ & $d^2$ & $\sin\theta$ \\ \hline\hline
  0.9999 & 0.000015 & 0.999998 & 3.94$\tt^{-9}$ & \bf 1 & 0.00016 \\ \hline
  0.8 & 0.0463 & 0.9951 & 4.07$\tt^{-9}$ & 3.22$\tt^{-4}$ & 0.00383 \\ \hline
  0.656 & 0.0697 & 0.9925 & 4.16$\tt^{-9}$ & 2.10$\tt^{-4}$ & \bf 0.00 \\ \hline
  0.6 & 0.0770 & 0.9916 & 4.19$\tt^{-9}$ & 1.88$\tt^{-4}$ & -0.00195 \\ \hline
  0.4 & 0.0959 & 0.9894 & 4.31$\tt^{-9}$ & 1.47$\tt^{-4}$ & -0.00992 \\ \hline
  0.2 & 0.1051 & 0.9883 & 4.43$\tt^{-9}$ & 1.31$\tt^{-4}$ & -0.01845 \\ \hline
  0.0 & 0.1062 & 0.9881 & 4.55$\tt^{-9}$ & 1.26$\tt^{-4}$ & -0.02649 \\ \hline
 -0.2 & 0.0997 & 0.9889 & 4.68$\tt^{-9}$ & 1.31$\tt^{-4}$ & -0.03314 \\ \hline
 -0.4 & 0.0860 & 0.9906 & 4.80$\tt^{-9}$ & 1.48$\tt^{-4}$ & -0.03744 \\ \hline
 -0.6 & 0.0650 & 0.9930 & 4.94$\tt^{-9}$ & 1.89$\tt^{-4}$ & -0.03806 \\ \hline
 -0.8 & 0.0365 & 0.9962 & 5.08$\tt^{-9}$ & 3.27$\tt^{-4}$ & -0.03239 \\ \hline
 -0.9999 & 0.000012 & 0.999998 & 5.23$\tt^{-9}$ & \bf 1 & -0.00064\\ \hline
\end{tabular}\\[1\baselineskip]
Table 1: Numerical result for the relevant parameters fixed by quark masses and Cabibbo angle.  
\end{figure}
 
Thus we have confirmed that observed quark masses and flavor mixing angle can be reproduced
in our model of GHU. 
Let us note that in eq.\,(\ref{Cabibbo}) $\theta_c$ disappears
in the limit of universal bulk mass, 
i.e. $M^1 = M^2$ and therefore $c = d$, as is expected.  

Some comments are in order. 
One might think that the above analysis of the diagonalization of fermion mass matrices restricting only to the zero mode sector is not complete, 
since it ignores possible mixings between zero mode and massive exotic states 
and the zero mode and non-zero KK modes given in section 2 may mix with each other to form mass eigenstates 
once the VEV $\langle A_y \rangle$ is switched on. 
(Let us recall that the standard model quark doublets do not mix with brane localized fermion by construction.) 
Such mixings, however, are easily known not to exist in the limit of vanishing VEV, $\langle A_y \rangle = 0$. 
Hence, even in the presence of the VEV such mixings will be suppressed by the small ratios of the VEV to the compactification scale 
or large brane localized masses. 
Therefore, our analysis is a good approximation at the leading order. 

Introducing the source of flavor mixing in the brane localized masses has already been considered in \cite{Csaki}, for instance. 
The difference between their model and ours is that 
in our model the interplay with the bulk masses and the Yukawa couplings in the bulk is crucial, while the flavor mixing is put by hand in ref.\,\cite{Csaki},
since Yukawa coupling is not allowed in the bulk in the model.

\section{$\bs{K^0-\bar K^0}$ mixing}
In this section, we apply the results of the previous section to a representative FCNC process 
due to the flavor mixing, $K^0-\bar{K}^0$ mixing, responsible for the mass difference of two neutral kaons $\Delta m_{K}$. 

As we have discussed in the introduction, in our model natural flavor conservation is not realized, i.e. 
FCNC processes are possible, already at the tree level. 
We restrict ourselves to the FCNC processes of zero mode down-type quarks due to gauge boson exchange at the tree level. 
First let us consider the processes where zero mode gauge bosons are exchanged. 
If such type of diagrams exist with a sizable amplitudes, it  may easily spoil the viability of the model. 
Concerning the zero mode gauge boson exchange, the gauge couplings are universal for fermions including non-zero KK modes, 
i.e. they are generation independent and depend just on the relevant quantum numbers such as $I_3$, 
since the mode function of the zero mode gauge boson is $y$-independent and 
the couplings become universal simultaneously with the normalization of kinetic term of fermions. 
Thus in this case the condition proposed by Glashow-Weinberg \cite{GW} is applicable straightforwardly. 
At the first glance, the condition of Glashow-Weinberg seems to be not satisfied in our model, 
since there are right-handed down-type quarks belonging to different representations, 
i.e. quarks belonging to $\psi ({\bf 3})$ and $\psi (\bar{\bf 6})$ of $SU(3)$. 
Then the Z boson exchange seems to yield FCNC. 
Fortunately, however, 
the down-type quarks belonging to $\psi (\bar{\bf 6})$ (more precisely $\Sigma^i$) is known to have the same quantum number $I_3$ 
as that of $d^i$ belonging to $\psi ({\bf 3})$, and FCNC does not arise even after moving to the mass eigenstates. 
The exchanges of zero mode photon and gluon trivially do not possess FCNC.           

Hence, the remaining possibility is the process with the exchange of non-zero KK gauge bosons. 
Now the mode functions of these gauge bosons are $y$-dependent and their couplings with fermions are no longer universal. 
Namely the different bulk mass of fermion for each generation is a new source of flavor violation 
and the coupling constants in the effective four-dimensional lagrangian become generation-dependent, 
thus leading to FCNC after moving to the mass eigenstates.  

Along this line of argument, we study $K^0-\bar K^0$ mixing caused by the non-zero KK gluon exchange at the tree level, as the dominant contribution to this FCNC process.

For such purpose, we derive the strong interaction vertices: restricting to the zero mode sector of down-type quarks 
and integrating over the fifth dimensional coordinate $y$, we obtain the relevant four-dimensional interactions: 
\begin{align}
	\label{strong}
	\mathcal{L}_\s
 \supset\,&
	\frac{g_\s}{2\sqrt{2\pi R}}G_\mu^a
	\Big(
	\bar d^i_R\gamma^\mu\lambda^ad^i_R
	+\bar Q^i_{3L}\lambda^a\gamma^\mu Q^i_{3L}+\bar Q^i_{6L}\lambda^a\gamma^\mu Q^{i}_{6L}
	\Big)\notag\\
 &  +\frac{g_\s}2G_{\mu}^{a(m)}\!
	\left\{
	\bar d_R^i\lambda^a\gm d_R^iI_{RR}^{i(0m0)}
	+(-1)^m\big(
	 \bar Q^i_{3L}\lambda^a\gm Q^i_{3L}
	 +\bar Q^i_{6L}\lambda^a\gm Q^i_{6L}
	 \big)
	I_{RR}^{i(0m0)}
	\right\}\notag\\
 \supset\,&
	\frac{g_\s}{2\sqrt{2\pi R}}G_\mu^a\!
	\left(
	\bar{\tilde d}_R^i\gm\lambda^a\tilde d_R^i
	+\bar{\tilde d}_L^i\lambda^a\gm \tilde d_L^i
	\right)\notag\\
 &  +\frac{g_\s}2G_\mu^{a(m)}
	\bar{\tilde d}_R^i\lambda^a\gm\tilde d_R^j\!
	\left(V_{dR}^\dag I_{RR}^{(0m0)}V_{dR}\right)_{ij}\notag\\
 &  +\frac{g_\s}2G_\mu^{a(m)}
	\bar{\tilde d}^i_L\lambda^a\gm\tilde d_L^j
	(-1)^m\!\left(
	V_{dL}^\dag U_3^\dag I_{RR}^{(0m0)}U_3V_{dL}
	+V_{dL}^\dag U_4^\dag I_{RR}^{(0m0)}U_4V_{dL}
	\right)_{ij}
\end{align}
where $I_{RR}^{(0m0)}$ is a matrix with the element $\delta_{ij}I_{RR}^{i(0m0)}$,
where ``vertex function"    
\begin{equation} 
\label{vertexfunction} 
   I_{RR}^{i(0m0)}
 = \frac1{\sqrt{\pi R}} \dy\!\big(f^{i}_{R}\big)^{2}\cos (M_m y)
 = \frac1{\sqrt{\pi R}}\frac{4(M^{i})^{2}}{4(M^{i})^{2} + \left(\frac{m}{R}\right)^{2}} 
   \frac{(-1)^{m}\e^{2\pi M^{i}R} - 1}{\e^{2\pi M^{i}R} - 1},    
\end{equation}
where the mode expansion of gluon
\begin{equation}
	G_\mu(x,y)
  = \frac1{\sqrt{2\pi R}}G_\mu
	+\sum_{n=1}^\infty\frac1{\sqrt{\pi R}}G_\mu^{(n)}\cos\left(M_ny \right)
\end{equation}
has been substituted.
Let us note that the vertex functions for left-handed quarks $I_{LL}^{i(0m0)}$ is related
to $I_{RR}^{i(0m0)}$ as $I_{LL}^{i(0m0)} = (-1)^mI_{RR}^{i(0m0)}$,
since the exchange of chiralities corresponds to the exchange of two fixed points. 
In eq.\,(\ref{strong}), $\tilde{d}$ denotes mass eigenstates,
$\big(\tilde d^1, \tilde d^2\big) = \big(\tilde d, \tilde s\big)$. 
The derivation of the last line of the equation is easily understood, 
since $Q_{3L} \sim U_{3} Q_{\SM L}, \ Q_{6L} \sim U_{4} Q_{\SM L}$ ignoring $Q_{\H L}$ 
and  $Q^{i}_{\SM L} = \big(u^{i}_L, d^{i}_L\big)$, $\big(d^{1}_L, d^{2}_L\big)^{t} = V_{dL}\big(\tilde{d}^{1}, \tilde{d}^{2}\big)^{t}$. 

We can see from (\ref{strong}) that the FCNC appears in the couplings of non-zero KK gluons 
due to the fact that $I_{RR}^{(0m0)}$ is not proportional to the unit matrix (the breaking of universality), 
while the coupling of the zero mode gluon is flavor conserving, as we expected.

\begin{figure}[h]
\begin{center}
\includegraphics{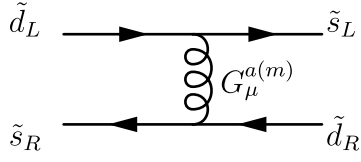}
\end{center}
\caption{The diagram of $K^0-\bar K^0$ mixing via KK gluon exchange}
\label{fig1}
\end{figure}

The Feynman rules necessary for the calculation of $K^0-\bar{K}^0$ mixing can be read off from (\ref{strong}).
\begin{align}
\begin{array}{c}
\includegraphics{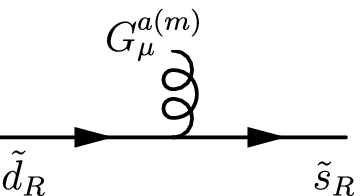}
\end{array}
&=
 \frac{g_\s}{2}\! \left(V_{dR}^\dag I_{RR}^{(0m0)}V_{dR}\right)_{21}\!\lambda^a\gm R\\[4pt]
\begin{array}{c}
\includegraphics{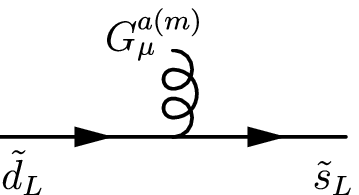}
\end{array}
&=
 \frac{g_\s}{2} (-1)^m\!
 \left(
 V_{dL}^\dag U_3^\dag I_{RR}^{(0m0)}U_3 V_{dL}
 +V_{dL}^\dag U_4^\dag I_{RR}^{(0m0)}U_4V_{dL}\right)_{21}\!\lambda^a \gm L\\
\begin{array}{c}
\includegraphics{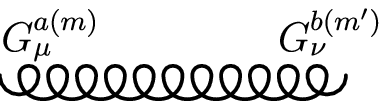}\!\!\!
\end{array}
&=
 \delta_{mm'}\delta_{ab}\frac{\eta_{\mu\nu}}{k^2-M_m^2}
 ~~~~{\Big(\,\text{'t Hooft-Feynman gauge}\,\Big)}
 \label{propagator}
\end{align}
The non-zero KK gluon exchange diagram, which gives the dominant contribution to the process of $K^0-\bar{K}^0$ mixing, is depicted in Fig.\,\ref{fig1}.
The reason why its contribution is dominant is that the diagram yields effective four-Fermi operator, 
which is product of left-handed and right-handed currents.  
Let us note that in the standard model the 1-loop box diagram yields four-Fermi operator, 
which is product of pure left-handed currents. 
Thus in the case of the standard model to form a pseudo-scalar state, the neutral $K$ meson, 
from $\tilde{d}$ and $\bar{\tilde{s}}$, chirality flip is needed and the amplitude is suppressed by small current quark masses. 
On the other hand, the four-Fermi operator of our interest has both left-handed and right-handed quarks 
and the amplitude is not suppressed by small quark masses. 
This means the amplitude is relatively enhanced compared to the case of the standard model 
with an ``enhancement" factor $\frac{m_K}{m_d + m_s}$, as we will see below.   

One may wonder whether the exchange of extra space component of gluon, $G_y^{a(m)}$, 
also gives similar contribution with the enhancement factor, 
since scalar-type coupling causes chirality flip. 
We, however, find the contribution is relatively suppressed 
by small masses of external quarks $m_q \ (m_q = m_d,\, m_s)$. 
Let us note that the zero mode of $G_y^{a(m)}$ ($m = 0$) is ``modded out" 
by orbifolding and non-zero KK modes of $G_y^{a(m)}$ ($m \neq 0$) are absorbed 
as the longitudinal components of massive gluons $G_\mu^{a(m)}$ through Higgs-like mechanism. 
In the unitarity gauge, the contribution of such longitudinal components are taken into account 
by adding to the propagator eq.(\ref{propagator}) a piece proportional to $\frac{k_{\mu}k_{\nu}}{M_{n}^{2}}$, 
where $k_{\mu}$ is the momentum transfer. 
By use of equations of motion for external quarks, 
its contribution to the amplitude is relatively suppressed by a factor $\frac{m_q^{2}}{M_{n}^{2}} = {\cal O} \big(m_q^{2}R^{2}\big)$ 
and we can safely neglect the contribution of $G_y^{a(m)}$ exchange. 

By noting the fact $k^2\ll M_n^2$ for $n \neq 0$, 
the contribution of diagram of Fig.\,\ref{fig1} is written in the form of effective four-Fermi lagrangian obtained by use of Feynman rules listed above,  
\begin{align}
	\begin{array}{c}
	 \includegraphics[scale=0.8]{KKgluonexchange.eps}
	\end{array}
 \sim&
	-\sum_{m=1}^\infty\frac{g_\s^2}4\frac{(-1)^m}{M_m^2}\!
	\left(
	V_{dL}^\dag U_3^\dag I_{RR}^{(0m0)}U_3V_{dL}
	+V_{dL}^\dag U_4^\dag I_{RR}^{(0m0)}U_4 V_{dL}
	\right)_{21}\notag\\
 &  \label{KKbar}
	\times\!\left(V_{dR}^\dag I_{RR}^{(0m0)}V_{dR}\right)_{21}\!
	\left(\bar{\tilde s}_L\lambda^a \gm \tilde d_L\right)\!\!
	\left(\bar{\tilde s}_R\lambda^a \gamma_\mu  \tilde d_R\right)\ .
\end{align} 
The sum over the integer $m$ is convergent
and the coefficient of the effective lagrangian (\ref{KKbar}) is suppressed by $1/M_c^{2}$,
where $M_c = 1/R$ is the compactification scale:
the decoupling effects of non-zero KK gluons.
We can verify utilizing the unitarity condition
that the coefficient vanishes in the limit of universal bulk masses $M^1=M^2= \cdots$
and therefore when $I_{RR}^{(0m0)}$ is proportional to the unit matrix, 
as we expect since in this limit flavor mixing just disappears; 
\begin{equation}
\left\{
\begin{aligned}
 &  V_{dL}^\dag\!
	\left(U_3^\dag I_{RR}^{(0m0)}U_3+U_4^\dag I_{RR}^{(0m0)}U_4\right)\!
	V_{dL}
	\xrightarrow{M^1=M^2= \cdots}
	V_{dL}^\dag\! \left(U_3^\dag U_3+U_4^\dag U_4\right)\! V_{dL}I_{RR}^{(0m0)}
	\propto {\mathbf 1}_{n\times n}\ ,\\
 &  V_{dR}^\dag I_{RR}^{(0m0)}V_{dR}
	\xrightarrow{M^1=M^2= \cdots} 
	V_{dR}V_{dR}^\dag I_{RR}^{(0m0)}
	\propto {\mathbf 1}_{n\times n}\ .
\end{aligned}
\right.
\end{equation}
The relevant hadronic matrix elements are written
by use of the ``bag parameters" $B_4,\, B_5$, 
which denote the deviation from the approximation of vacuum saturation
and whose numerical results are obtained
by lattice calculations $B_4=0.81$, $B_5=0.56$ \cite{BBDGW}: 
\begin{eqnarray}
 \big\langle \bar K\big|\bar s_\alpha \gamma^\mu L d_\alpha \cdot \bar s_\beta \gamma_\mu R d_\beta \big|K\big\rangle
&\!\approx\!&
 \frac{B_5}{6}\left(\frac{m_K}{m_d+m_s}\right)^{\!2}\!f_K^2 m_K,
 \\
 \big\langle \bar K \big|\bar s_\alpha \gamma^\mu L d_\beta \cdot \bar s_\beta \gamma_\mu R d_\alpha \big|K\big\rangle
&\!\approx\!&
 \frac{B_4}{2}\left(\frac{m_K}{m_d+m_s}\right)^{\!2}\!f_K^2 m_K,   
\end{eqnarray}
where $\alpha, \beta$ are color indices and $f_K(\simeq 1.23f_\pi)$ is the kaon decay constant. 
$m_K,\,m_d,\,m_s$ denote the kaon mass and the current quark masses of down and strange quarks.  
Note that the color indices are contracted in different ways in these two matrix elements. 
Only the terms with the chiral enhancement factor are left in the above expressions.  
Using these results, the hadronic matrix element of the effective four-Fermi operator is obtained as  
\begin{equation}
\begin{aligned}
&\frac14\big\langle \bar K \big|\bar s \lambda^a \gamma^\mu Ld\cdot \bar s \lambda ^a \gamma_\mu R d\big|K\big\rangle \nt\\
&=
-\frac{1}{6} \big\langle \bar K \big| \bar s_{\alpha L} \gm  d_\alpha \cdot
 \bar s_{\beta R}\gamma_\mu d_{\beta R} \big| K \big\rangle
 +\frac{1}{2} \big\langle \bar K \big|\bar s_{\alpha L}\gm d_{\beta L}\cdot 
 \bar s_{\beta R}\gamma_\mu d_{\alpha R} \big| K \big\rangle \nt\\
&\approx
 \left(\frac{B_4}{4}-\frac{B_5}{36}\right)\!\!
 \left(\frac{m_K}{m_d+m_s}\right)^{\!2}\! f_K^2 m_K. 
\end{aligned}
\end{equation}
Thus the contribution of KK gluon exchange to $K_L -K_S$ mass difference is given as 
\begin{eqnarray}
 \Delta m_K(\text{KK})
&\!\!\!=\!\!\!&
 2 \big\langle \bar K\big|\mathcal {L}^{\Delta S=2}_{\rm eff}\big|K\big\rangle
\nt\\
&\!\!\!\approx\!\!\!&
 -2\sum_{m=1}^\infty \frac{g_\s ^2}{M_m^2}(-1)^m\!
   \left(\frac{B_4}{4}-\frac{B_5}{36}\right)\!\!\left(\frac{m_K}{m_d+m_s}\right)^{\!2}\!f_K^2 m_K \nt\\
&&\label{massdiff}
 \times\!\left(V_{dL}^\dag U_3^\dag I_{RR}^{(0m0)}U_3V_{dL}+V_{dL}^\dag U_4^\dag I_{RR}^{(0m0)}U_4 V_{dL}\right)_{21}\!\!
 \left(V_{dR}^\dag I_{RR}^{(0m0)}V_{dR}\right)_{21}\!.
\end{eqnarray}

So far the obtained results for $K^{0} - \bar{K}^{0}$ mixing and $\Delta m_K(\text{KK})$ are valid for an arbitrary number $n$ of generations. 
{}From now on we focus on the simplified two generation scheme in order to estimate the mass difference. 
It would be more desirable to discuss the FCNC process in the full three generation scheme. 
We, however, realize that in the standard model the first two generations give important contributions 
to the CP conserving observable $\Delta m_K$. So, we expect that our analysis gives a reasonable result. 
We thus obtain 
\begin{equation} 
\label{massdiff1}
   \Delta m_K(\text{KK})
 \approx 
   -16 \pi \alpha_\s R^2 A \cdot  
   \pi R \sum_{m=1}^\infty \frac{(-1)^m\!}{m^2}\!\left(I_{RR}^{1(0m0)} - I_{RR}^{2(0m0)}\right)^{\!2}\!
   \left(\frac{B_4}{4}-\frac{B_5}{36}\right)\!\!\left(\frac{m_K}{m_d+m_s}\right)^{\!\!2}\!\!f_K^2 m_K \nt
\end{equation} 
where the ordinary four-dimensional $\alpha_\s$ is defined by
$\alpha_\s = \frac{(g_\s^{4D})^2}{4\pi}=\frac{1}{2\pi R}\frac{g_\s^2}{4\pi}$.
The coefficient $A$ in (\ref{massdiff1}) is defined as follows:
\begin{eqnarray}
& A \equiv &\!\!\!
 \left\{
 V_{dL}^\dag\!\left(U_3^\dag \frac{\sigma_3}2 U_3+U_4^\dag \frac{\sigma_3}2 U_4\right)\!V_{dL}
 \right\}\!
 \left(V_{dR}^\dag \frac{\sigma_{3}}{2} V_{dR}\right)\!
 = \sin \theta_d \cos \theta_d (\alpha + \alpha'), \nonumber \\
& \alpha \equiv &\!\!\!
 -\frac{1}{2}\big(1-a^2\big) \sin 2\theta_{dL} \cos^2 \theta
 +\frac{1}{2}\big(1-b^2\big) \sin 2\theta_{dL} \sin^2 \theta  \nonumber \\ 
&&\!\!\!
 -\frac{1}{2} \sqrt{(1-a^2)(1-b^2)} \cos 2\theta_{dL} \sin 2 \theta, \nonumber \\ 
& \alpha' \equiv &\!\!\!
 -\frac{1}{2} a^2 \sin 2\theta_{dL} \cos^2 \theta' + \frac{1}{2} b^2 \sin 2\theta_{dL} \sin^2 \theta'
 -\frac{1}{2} ab \cos 2\theta_{dL} \sin 2 \theta',    
\label{massdiff2}
\end{eqnarray}
where $\sigma_{3}$ is one of Pauli matrices and $\theta_d$ is an angle in the rotation matrix $V_{dR}$ to diagonalize $I_{RL}^{(00)}U_3 U_3^\dag I_{RL}^{(00)}$:  
\begin{equation}
   \tan 2\theta_d 
 = \frac{2(b^2-a^2) cd \sin \theta \cos \theta} 
		{c^2 \big\{(1-a^2) \cos^2 \theta + (1-b^2) \sin^2\theta \big\}
		-d^2 \big\{(1-a^2) \sin^2 \theta + (1-b^2) \cos^2\theta \big\}}\ .\notag 
\end{equation}

The constant $\alpha_{\s}$ should be estimated at the scale $\mu_K = 2.0$\,GeV
where the $\Delta S=2$ process takes place.
The reason is that the bag parameters in \cite{BBDGW} are those estimated
at the scale of $\mu_K$. 
So we have to take into account the renormalization group effect from the weak scale, 
where $\alpha_\s$ is known rather precisely, down to $\mu_K$:
\begin{align} 
	\alpha_\s^{-1}(\mu_K)
  = \alpha_\s^{-1}(m_Z)
	+\frac1{6\pi}\!
	\left(
	23\ln\frac{m_Z}{m_b}+25\ln\frac{m_b}{\mu_K}
	\right)
  \quad \longrightarrow \quad
  \alpha_\s(\mu_K) \approx 0.268
\end{align}
where $\alpha_\s(m_Z) \approx 0.118$ has been put. 

Combining these results, we finally obtain 
\begin{equation} 
\label{deltamfinal}
	\Delta m_K(\text{KK})
 \simeq
	2.09\times10^4\cdot(R f_{\pi})^2\sin2\theta_d(\alpha + \alpha')\,
	\pi R\sum_{m=1}^\infty
	\frac{(-1)^m\!}{m^2}\!
	\left(I_{RR}^{1(0m0)}-I_{RR}^{2(0m0)}\right)^{\!2}\,[\mbox{MeV}].
\end{equation}
The room for the \lq\lq New Physics\rq\rq contribution $\Delta m_K(\text{NP})$ is given
by the difference between the experimental data and the standard model prediction
\cite{IL}, \cite{BS}: 
\begin{align}
	\big|\Delta m_K(\text{NP})\big|
 &= \big|\Delta m_K(\text{Exp})-\Delta m_K(\text{SM})\big|
  = \left|1-\left(\frac76\sim\frac56\right)\right| \Delta m_K(\text{Exp})\notag\\
 &< \frac16\Delta m_K({\rm Exp})
  = \frac16\cdot3.48\times10^{-12}~[{\rm MeV}]\ .
\end{align}
Identifying $\Delta m_K(\text{NP})$ with our result $\Delta m_K(\text{KK})$, we obtain a lower bound for the compactification scale. Namely,  
\begin{align}
	\frac{1}{R}
   > 1.77\times10^4\sqrt{\left|\sin2\theta_d(\alpha+\alpha')\,
	\pi R\sum_{m=1}^\infty
	\frac{(-1)^m\!}{m^2}\!
	\left(I_{RR}^{1(0m0)}-I_{RR}^{2(0m0)}\right)^{\!2}\right|}~[{\rm TeV}]\ .
\end{align} 

\begin{figure}[h]
\centering
\includegraphics[scale=.45]{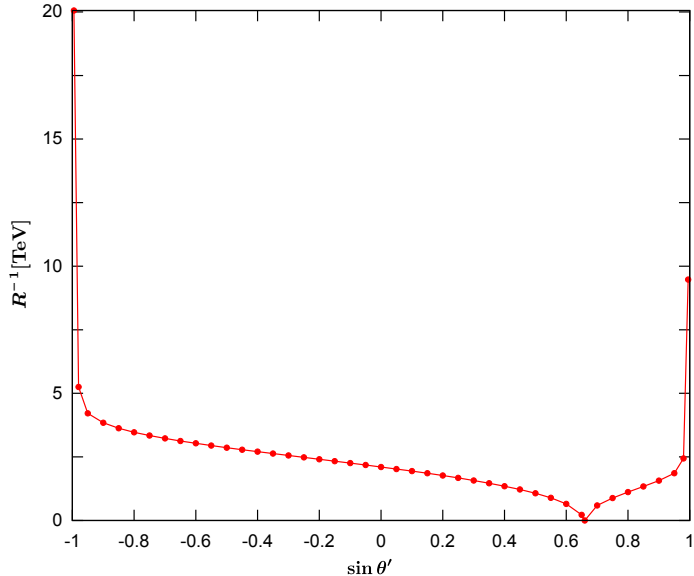}
\caption{The lower bound on $\frac1R$ as a function of $\sin\theta'$.}
\label{fig2}
\end{figure}

Since our theory has one unfixed parameter, say $\theta'$,
the lower bound depends on it.
The obtained result is given in Fig.\,\ref{fig2},
where the lower bound on $R^{-1}$ is plotted as a function of $\theta'$.
If we demand that the prediction of our model is consistent with the data on $\Delta m_K$,
irrespectively of the choice of $\theta'$,
we should require that compactification scale is greater
than the possible largest value in Fig.\,\ref{fig2}, $R^{-1} \gtrsim 20\,{\rm TeV}$.
We, however, also would like to point out that in most cases,
except for the extreme case $|\sin \theta'| \simeq 1$,
the obtained lower bound is rather mild,
e.g. $2.1 \ {\rm TeV}$ for $\theta' = 0$.
As we discuss below, this is a genuine feature of GHU scenario. 

A few comments are in order.
Fig.\,\ref{fig2} indicates that the lower bound becomes 0
for a special case $\sin \theta' \simeq 0.66$.
What is happening in this situation is that $\theta = \theta_{dL} = \theta_d = 0$
and the contribution of KK gluon exchange accidentally vanishes.
As the matter of fact,
in this specific case the 4-Fermi interaction
with the product of currents with the same chirality, L-L or R-R,
induced by the KK gluon exchange becomes dominant contribution.
The allowed lower bound on $R^{-1}$ due to this kind of processes is expected to be mild,
since the hadronic matrix elements of the 4-Fermi operators have no chiral enhancement factor
and therefore the obtained bound is expected to be smaller
by factor $\frac{m_d+m_s}{m_K} \simeq 0.21$
compared to the bound obtained in the case of 4-Fermi operator of the type L-R,
considered in this paper.
Even if we take the most stringent lower bound $20\,{\rm TeV}$
in the case of L-R as a reference value,  
the corresponding bound in the case of the same chirality will be $4\,{\rm TeV}$.

Another comment is that,
let us emphasize the presence of ``GIM-like" mechanism in GHU scenario, which turns out to exist in the tree level diagram due to KK gluon exchange. Note that the lower bound obtained above is smaller (except the extreme cases of $|\sin \theta'| \simeq 1$) than what we naively expect assuming that the vertex of the tree level diagram relevant for the FCNC process is of the order 
${\cal O} (\sin \theta_{c} \cos \theta_{c})$. Namely, if we write the Wilson coefficient of the four-Fermi operator (the ordinary $V-A$ type is assumed ) generically in a form 
\begin{align}
\frac{(\sin \theta_{c} \cos \theta_{c})^{2}}{M^{2}},  
\end{align}
the lower bound for the mass scale is obtained by comparing with the data:     
\begin{align}
	\frac1{M^2} \leq 10^{-5}~\big[{\rm TeV}^{-2}\big]
\quad \longrightarrow \quad
M \geq 300 \,[\mbox{TeV}]\ ,
\end{align}
which is much larger than the lower bound we obtained in most cases of the choice of $\theta'$. 

This apparent discrepancy may be attributed to the GIM-like mechanism in our scenario, which we will see now. 
As has been already pointed out, the genuine feature of GHU scenario is that the non-universal bulk masses is a new source of flavor violation and FCNC processes at tree level is handled by the difference of bulk masses. In fact, the rate of $K^{0}-\bar{K}^{0}$ mixing is handled by the factor $\big(I_{RR}^{1(0m0)} - I_{RR}^{2(0m0)}\big)^{2}$ as is seen in (\ref{deltamfinal}), which is due to the non-universality of gauge coupling of KK gluons coming from the difference of bulk masses \big($M^1 \neq M^2$\big). We easily see that this factor is automatically suppressed for generations with light quarks, such as 1st and 2nd generations. 

Recall that in GHU hierarchical small quark masses are naturally realized without fine tuning
by exponential suppression factors $\e^{-\pi R M^i}$ coming from $I_{RL}^{i(00)}$.
On the other hand, when $\pi R M^i \gg 1$,
the ``width" $1/M^i$ of the mode function of localized zero-mode fermion is much smaller
than the period $\frac{2\pi R}m$
of mode function $\cos\!\big(\frac mRy\big)$ of KK gauge bosons.
(Note that smaller KK modes $m$ play potentially important role
in the convergent mode sum $\sum_m$.)
Then the exponential dumping of the fermion mode functions is so fast
that the mode functions of KK gauge bosons are effectively almost $y$-independent.
Thus the situation mimics that
in the case of zero-mode gauge boson and the gauge coupling of KK gauge bosons
becomes almost universal: in the limit $\pi RM^i \to \infty$,
as is seen from (\ref{vertexfunction}),
$I_{RR}^{i(0m0)}$ approaches to $\frac{(-1)^m}{\sqrt{\pi R}}$, a constant.
Therefore FCNC processes at the tree level should be automatically suppressed
for the processes of light quarks.
The suppression mechanism is similar to the famous GIM-mechanism
where FCNC is suppressed by a typical factor $\frac{m_c^2-m_u^2}{m_W^2}$.
This is the reason we call the suppression mechanism ``GIM-like" mechanism.
In fact, the KK mode sum 
\begin{equation} 
	S_{\rm KK}
 \equiv
	\pi R\sum_{m=1}^\infty
	\frac{(-1)^m}{m^2}\!
	\left(I_{RR}^{1(0m0)}-I_{RR}^{2(0m0)}\right)^{\!2}
\end{equation}
can be performed analytically,
though we do not present the result here.
When $\pi R M^i \gg 1$, an approximated formula is given as  
\begin{align}
	S_{\rm KK}
 \simeq
	-\frac{\pi^2}2\!\left(\e^{-2\pi M^1R}+\e^{-2\pi M^2R}\right)
	-\frac{\pi}{2R}\frac{(M^1)^2-M^1M^2+(M^2)^2\!}{M^1M^2(M^1-M^2)}\!
	\left(\e^{-2\pi M^1R}-\e^{-2\pi M^2R}\right).
\end{align}
Now we clearly see that there appear exponential suppression factors
$\e^{-2\pi M^1R}, \ \e^{-2\pi M^{2}R}$.
For instance, in the case of $\theta' = 0$,
$S_{\rm KK} \simeq -1.17 \times 10^{-5}$,
which is in very good agreement with
what we obtain by direct numerical calculation of the KK mode sum.
It is quite interesting to note that these suppression factors just correspond to the suppression factors in the Yukawa couplings and therefore to the squared ratios of small quark masses $m_{q}$ to $m_{W}$, $\big(\frac{m_{q}}{m_W}\big)^{2}$. Thus our suppression mechanism is similar to that of GIM mechanism. 

In the exceptional extreme case of $|\sin \theta'| \simeq 1$, the bulk mass $M^{2}$ happens to be relatively small and the suppression mechanism does not work. That is why we get severe lower bound on the compactification scale in the extreme cases. 

Finally, it is interesting to note that there seems to be some similarity of our suppression mechanism of FCNC with the GIM-like mechanism in the RS warped extra dimension model, where the lower bound of the compactification scale obtained from the analysis of FCNC is $2-30$ (TeV) \cite{GP}. 
In the scenario of warped extra dimension, it is claimed that the mode functions of KK gauge bosons are approximately flat near the UV (Planck) brane, where
quarks are localized, and KK gauge boson coupling to fermion is almost universal.

\section{Summary}
In this paper, we discussed flavor mixing and resultant flavor changing neutral current processes 
in the $SU(3) \otimes SU(3)_\text{color}$ gauge-Higgs unification scenario. 
These are caused by the violation of global flavor symmetry, ``flavor violation". 
To achieve flavor violation in the gauge-Higgs unification is a challenging issue, 
since the Yukawa couplings are originally given by (higher dimensional) gauge interaction, 
which is universal, i.e. generation-independent. 
Thus the question of whether the flavor mixing is realized or not is crucial for the viability of the scenario. 

We argued that there exists a new source of the flavor violation, i.e. the presence of $Z_2$-odd bulk masses, 
which is a genuine feature of higher dimensional gauge theories with orbifold compactification, 
such as the models of gauge-Higgs unification. 
It is, however, noticed that in the gauge-Higgs unification scenario 
the bulk masses can be made ``diagonal" in the base of generation by a suitable unitary transformation 
and are not enough to achieve flavor mixing, though they succeed in explaining the hierarchical small fermion masses in a natural way without fine tuning by the exponential suppression factor due to the localization of the mode functions of fermion zero modes. 
We are thus led to introduce brane-localized masses, 
which are necessary to remove redundant $SU(2)$ quark doublet from the zero mode sector anyway in our model. 
We have explicitly shown in a simplified two generation framework that under the presence of brane-localized masses, 
which can be ``off-diagonal" in the base of generation, 
the Yukawa couplings of Higgs field, which are originally generation-independent, 
recover the observed Cabbibo mixing together with quark masses.          

As a remarkable feature of the gauge-Higgs unification scenario, 
we have found that not only the presence of the brane-localized masses but also the interplay with the bulk masses is essential to get the flavor mixing, i.e. when the flavor violation due to the bulk masses goes away, 
or equivalently in the limit of degenerate bulk masses, flavor mixings exactly disappear for arbitrary number of generations. 
It should be emphasized that this is true even though the brane-localized mass parameters can be arbitrary, 
and that this property is the remarkable property of our scenario, 
not shared by, e.g., the scenario of universal extra dimension, 
where flavor mixing may be realized by ordinary Yukawa couplings without the need of bulk masses.      

Having flavor mixings, we discussed the resultant flavor changing neutral current (FCNC) processes, 
which have been playing central roles for the check of viability of new physics models.  
Especially an important question was raised of whether natural flavor conservation is realized in the gauge-Higgs unification scenario, 
i.e. whether FCNC processes are forbidden at the tree level. 
It has been realized that even though the condition proposed by Glashow and Weinberg \cite{GW} is satisfied,  the new source of flavor violation, i.e. non-degenerate bulk masses, leads to non-universal coupling constants of non-zero KK modes of gauge bosons with quarks. 
Such couplings then yield FCNC processes of quarks at tree level 
through flavor mixings due to the exchange of non-zero KK gauge bosons. 
Once again, we can confirm that in the limit of degenerate bulk masses the FCNC processes just disappear. 
       
As a typical example of FCNC processes under the precision tests, 
we calculated the prediction of our model concerning the effective lagrangian for $K^0 - \bar{K}^0$ mixing 
due to the non-zero KK gluon exchange at the tree level as the dominant contribution. 
Although the process is caused at the tree level, 
the obtained result for the mass difference of neutral kaon $\Delta m_K$ is finite 
and is suppressed by the inverse powers of compactification scale 
(the decoupling effects of heavy non-zero KK gluons) and may be handled as far as the compactification scale is high enough. 

Identifying the prediction with the room left for the New Physics contribution, 
i.e. the difference between the data and the standard model prediction, 
we have obtained the lower bound of the compactification scale. Since our theory has one unfixed parameter, say $\theta'$, the lower bound has been obtained as a function of the unknown parameter (see Fig.\ref{fig2}). 
We have found that except some extreme cases $|\sin \theta'| \simeq 1$, in the most range of the unknown parameter the lower bound is rather mild, say ${\cal O} (10) {\rm TeV}$, which is considerably smaller than we naively expect by assuming Cabibbo mixing. 

We have argued that the reason to get such smaller lower bound on the compactification scale is the presence of 
``GIM-like" mechanism in GHU scenario. Namely, in GHU scenario hierarchical small quark masses are naturally realized by exponential suppression factors $\e^{- \pi R M^{i}} \ (M^{i}: {\rm bulk \ mass})$. On the other hand, when $\pi R M^{i} \gg 1$, the ``width" $1/M^{i}$ of the mode function of localized zero-mode fermion is so small compared with the period  $\sim 2\pi R$ of mode function of KK gauge bosons that the mode functions of KK gauge bosons are effectively almost $y$-independent.
Thus the situation mimics that in the case of zero-mode gauge boson and gauge couplings of KK gauge bosons become almost universal and therefore FCNC processes at the tree level is automatically suppressed for the processes of light quarks. We have confirmed that the suppression is due to the exponential  suppression factor $\e^{- 2\pi R M^{i}}$, the same factor as the one to explain small quark masses, and the suppression factor is proportional to squared ratio of small quark masses to weak scale, $\frac{m_{q}^{2}}{m_{W}^{2}}$. Thus our suppression mechanism of FCNC is similar to the famous GIM-mechanism where FCNC is suppressed typically by a factor $\frac{m_{c}^{2} - m_{u}^{2}}{m_{W}^{2}}$. 

\subsection*{Acknowledgments}

This work was supported in part by the Grant-in-Aid for Scientific Research 
of the Ministry of Education, Science and Culture, No. 18204024 and No. 20025005.



\end{document}